\documentclass[final,5p,times,twocolumn, number,sort&compress]{elsarticle}

\DeclareGraphicsExtensions{.pdf}

\usepackage{amssymb}
\usepackage[pdfauthor={Thomas Jaouen}]{hyperref}

\newcommand{\degree}{\ensuremath{^{\circ}}}

\journal{Journal of Electron Spectroscopy and Related Phenomena}

\begin{document}

\begin{frontmatter}

\title{Photoelectron diffraction of twisted bilayer graphene}

\author[inst1]{S. Tricot}
\author[inst2]{H. Ikeda}
\author[inst1]{H.-C. Tchouekem}
\author[inst1]{J.-C. Le Breton}
\author[inst1]{S. Yasuda}
\author[inst2]{P. Kr\"uger}
\author[inst1]{P.  Le F\`evre}
\author[inst1]{D. S\'ebilleau}
\author[inst1]{T. Jaouen\corref{cor1}}
\ead{thomas.jaouen@univ-rennes.fr}
\cortext[cor1]{Corresponding author}
\author[inst1]{P. Schieffer}

\affiliation[inst1]{organization={Univ Rennes, CNRS, IPR (Institut de Physique de Rennes) - UMR 6251}, postcode={F-35000}, city={Rennes}, country={France}}
\affiliation[inst2]{organization={Graduate School of Engineering and Molecular Chirality Research Center, Chiba University}, city={Chiba}, postcode={263-8522}, country={Japan}}

\begin{abstract}
Photoelectron diffraction (PED) is a powerful spectroscopic technique that
combines elemental resolution with a high sensitivity to the local atomic
arrangement at crystal surfaces, thus providing unique fingerprints of selected
atomic sites in matter. Stimulated by the rapid innovation in the development
of various analysis methods for probing the atomic and electronic structures of
van der Waals (vdW) heterostructures of two-dimensional materials, we present a
theoretical assessment of the capacity of PED for extracting structural
properties such as stacking, twist angles and interlayer distances. We provide
a complete description of PED for the benchmark vdW heterostructure bilayer
graphene (BLG), by calculating and analyzing the PED of BLG in Bernal and
AA-stacking as well as twisted BLG for a wide range of the twist angle.
\end{abstract}


\end{frontmatter}

\section{Introduction}

In the past few years, the remarkable progress in controlling the stacking of
atomic sheets in van der Waals (vdW) heterostructures \cite{Kim2016,
Ribeiro2018} has opened up new avenues for manipulating electronic properties
by moiré superlattices. In two-dimensional (2D) materials, a moiré superlattice
can be formed by vertically stacking two layered materials with a twist angle
that generally modifies the electronic band structure through umklapp
scattering. At so-called “magic angle twist”, low-energy flat subbands appear
and electron interactions become the dominant energy scale, which may lead to
emergent electronic phases, such as correlated insulators, superconductors,
magnetism and topological electronic structures. In such a context, bilayer
graphene (BLG) appears as the simplest vdW heterostructure that has
demonstrated to exhibit correlated phases emerging from isolated flat bands
upon small-angle twisting \cite{Cao2018a, Cao2018b, Yankowitz2019, Sharpe2019,
Lu2019, Serlin2020}.

More recently, the observation of superconductivity and the quantum anomalous
Hall effect in Bernal bilayers \cite{Zhou2022, Geisenhof2021} and the discovery
of dodecagonal quasicrystalline order coexisting with Dirac fermions and
enhancing interlayer coupling in 30$^{\circ}$-twisted BLG \cite{Ahn2018,
Yao2018}, showed that large twist angles BLG constitute a platform for
exhibiting exotic physics as well. Whereas the moiré pattern of the dodecagonal
quasicrystal is purely incommensurate with 12-fold rotational symmetry and no
translational one, at twist angles satisfying the commensuration condition
$\theta_{t}=\arccos{\frac{3q^2-p^2}{3q^2+p^2}}$ with $p$ and $q$ integers,
twisted BLG forms spatially-periodic moiré pattern with a well-defined
elementary unit cell. While the values $\theta_{t} = 0^{\circ}$ and $\theta_{t}
= 60^{\circ}$ lead to the smallest possible unit cells with AA and Bernal (AB)
stacking, twisted BLG with commensurate rotation angles $\theta_{t}$ and
$60^{\circ}-\theta_{t}$ appear as pairs leading to unit cells of equal areas
but exhibiting distinct parities upon sublattice exchange (SE) \cite{Mele2010,
Mele2012}. If the commensuration unit cell contains A and B sublattice sites in
each layer coincident with atomic sites in the neigbouring layer, the SE parity
is even with points having six-fold symmetry. It is odd with three-fold
symmetry if only one sublattice site in the commensuration cell is eclipsed
\cite{Mele2012}. SE-even and SE-odd structures of a given commensuration pair
are related by inter-layer lattice translation and can be deduced from the
soustraction $q-p$ of the translation indices $(p,q)$ that is a factor of 3 for
even SE, only. Turning to the reciprocal space, the commensurate partners
exhibit electronic band structures having the same symmetry as Bernal and
AA-stacked BLG but with an interlayer coherence energy scale in the terahertz
regime  \cite{Mele2023a, Mele2023b}. As their untwisted parents, the coherent
scattering between the Dirac cones of the two layers gives rise to an interface
electronic structure exhibiting either massive Dirac bands, or a band gap
supporting topological edge states \cite{Kindermann2015, Koren2016}, depending
on the SE parity. 

To date, even if the increased sample quality hints at the possibility to
experimentally emcompass the extremely rich physics of BLG from
strongly-correlated electron phenomena at small twisted angles to
quasicrystalline order and interlayer coherent coupling at larger ones, only
few analysis methods allow for accurately accessing all together, the
structural, physical and chemical properties of real stacking. In this context,
photoelectron diffraction (PED) which exploits angle-resolved, core-level
photoemission from a crystal surface, appears as very promising since it is
element-specific and probes the \textit{local} atomic structure around the
emitting atoms at the \AA\ length scale. In that paper, we examine the
capability of PED for accessing local structural properties such as stacking,
twist angles and interlayer distances of BLG by a comprehensive set of
full-multiple scattering calculations. We first validate our theoretical
procedure by confronting calculated PED patterns to existing experimental
literature and demonstrate how sensitive PED is to the stacking geometry of
BLG. Motivated by the experimentally observed layer-by-layer resolution of the
C 1$s$ core level for BLG when grown on SiC(0001) \cite{Riedl2010, Razado2018},
or transferred on SiO$_2$ \cite{Kim2008, deLima2013, Matsui2013}, we also
extract the PED stereographic plots for the bottom and top layers of BLG. We
show that while the bottom layer PED provides information on the SE parity of
the stacking, the top layer PED is similar to the one of single-layer graphene.
This offers the unprecedented opportunity to determine the relative orientation
of two stacked graphene layers and stacking arrangements of moiré patterns once
twist angles are introduced. We then focus on twisted BLG for a wide range of
rotation angles between the magic-angle and that of the dodecagonal
quasicrystal ones, passing through various commensurate moiré crystals. We
unveil three distinct regimes of PED associated with small, intermediate and
large twisted-angles reflecting respectively the characteristic local Bernal
and AA stacking of large moiré supercells, the SE-parities of the small-unit
cell commensurate structures, and the 12-fold Stampfli tiling of the
quasicrystal. Finally, through the energy-dependence of the normal-emission
modulation function \cite{Tricot2022}, we end up by illustrating the high
sensitivity of PED to small variations of interlayer distances for both
primitive and twisted BLG.

\section{Computational details}

Full multiple scattering spherical wave cluster calculations have been
performed by using the MsSpec program \cite{Sebilleau2006,
Sebilleau2011,wwwMsSpec}. The phase shifts have been calculated using the
Hedin-Lundqvist exchange-correlation potential \cite{Hedin1970,
Hedin1971,Fujikawa2000} whose imaginary part describes the finite photoelectron
mean free path in the final state.  The inner potential value in-between
scattering centers was set to 12.5~eV. The polar angle ($\theta$) is defined
with respect to the surface normal and the azimuth angle ($\phi$) with respect
to the emission plane. The unpolarized photon source makes an angle of
45\degree\ with the electron analyzer direction. Both are fixed in the
laboratory frame while the sample is rotated in ($\theta, \phi$). Lattice
vibrations were described by averaging over $T$-matrix elements and using
isotropic mean-square displacements fixed to $5\times 10^{-3}$~\AA$^2$
\cite{Kuttel1994}. In the calculated PED stereographic projections, the
experimental resolution of the detector has been modeled by averaging
photoelectron intensities over 13 directions evenly distributed in an
acceptance cone with an angular aperture of 1.5\degree.

For each $(\theta,\phi)$ direction, the total PED intensity results from the
incoherent sum of the intensities associated with each inequivalent
photoelectron emitter in the structure. In AA-, Bernal-stacked or twisted BLG,
inequivalent emitters correspond to all carbon atoms contained in the
supercell. A standalone multiple scattering calculation is run for each
emitter, considering a set of approximately 400~atoms contained in a cylinder
of 13~\AA\ radius surrounding the emitter located on the cylinder's axis
oriented along the [001] direction of the twisted BLG. This cluster geometry
was carefully determined based on the convergence study with respect to the
system size and the memory requirements needed for the calculation. Here we
perform full multiple scattering calculations by matrix inversion rather than
using a series expansion. This has the great advantage of solving the multiple
scattering problem exactly. The downside is that the calculation requires a
large amount of RAM which increases quickly with the number of atoms in the
cluster $N$, and the maximum angular momentum $l_\textrm{max}$ used to expand
the photoelectron wavefunction in the spherical basis. The storage requirement
is proportional to $[N \times (l_\textrm{max} + 1)^2]^2$ with $l_\textrm{max}
\propto E^{1/2} r$ for kinetic energy $E$ and atomic radius $r$. It is worth
mentioning that the small size of a carbon atom makes it possible to keep
$l_\textrm{max}$ below a reasonable limit ($<15$), enabling PED patterns to be
computed in \textit{full}-multiple scattering at energies as high as 650 eV.
Since commensurate supercells considered in this study can contain as much as
11164~atoms, both process-based parallelism and standard shared memory
loop-based parallelism of Lapack routines were used to make the most of
computing power.

\section{Results and discussion}

\subsection{PED sensitivity to BLG stacking: Experiment vs. full-multiple scattering simulations.}

\begin{figure}[ht!]
\includegraphics[width=0.45\textwidth]{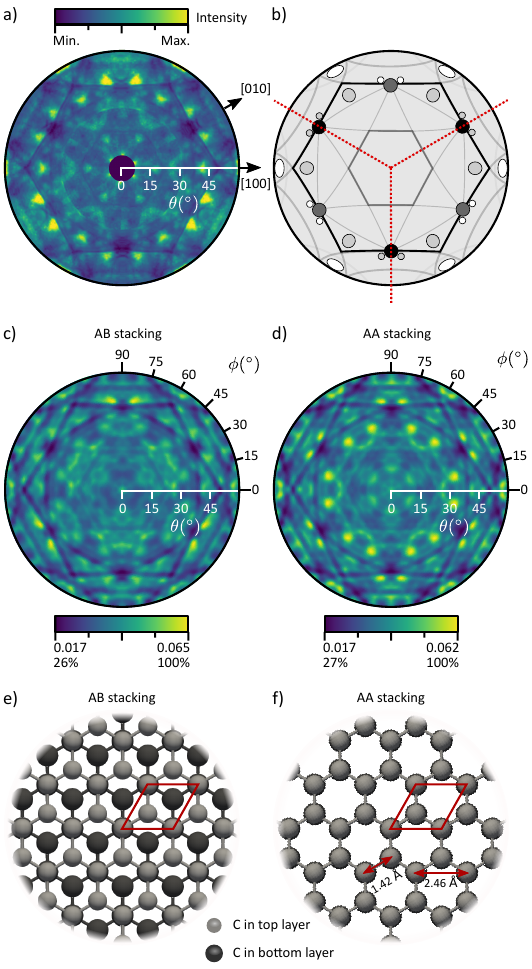}
\caption{
    [color online] a) Stereographic projection of PED data of the C 1$s$ core
    level at 638.6~eV kinetic energy (taken from~\cite{Razado2018}). b) Sketch
    of the main features found in the diffraction pattern. Red dashed lines
    highlight the three-fold symmetry. c)-d) Full-multiple scattering
    calculations of the stereographic projection of PED from C 1$s$ at 638.6 eV
    kinetic energy for a Bernal(AB)- c) and AA-stacked d) BLG. The calculations
    used the same experimental geometry as a). e)-f) Top views of the two
    superimposed C honeycomb lattices for Bernal (AB) e), and AA-stacked BLG
    f). The primitive cells of the Bernal and AA stacking having either half,
    or all of the sublattices sites eclipsed are highlighted in red on e) and
    f). Are also indicated on f) the C-C nearest neighbour distance
    ($1.42$~\AA) and the length of the basis vectors of the graphene unit cell
    ($2.46$~\AA).}
\label{fig1}
\end{figure}

In order to demonstrate how powerful full-multiple scattering simulations are
for accurately describing BLG, we first confront, in Fig. \ref{fig1},
calculated and experimental BLG PED data in stereographic projection. The
experimental data are taken from Ref.\cite{Razado2018} and were obtained for
BLG grown on SiC(0001) by recording the photoemission intensity of the C 1$s$
core level excited with a photon energy such that the kinetic energy of
photoelectrons was 638.6~eV [Fig. \ref{fig1}a)]. The experimental PED data
shows an overall hexagonal pattern and three-fold symmetry. The reduction from
six- to threefold symmetry is evident both on the ring of intense spots at
45$^{\circ}$ [small white and light-grey circles on Fig. \ref{fig1}b)], and on
the dark spots at $\sim$40$^{\circ}$ polar angles [large dark grey and dark
circles on Fig. \ref{fig1}b)]. 

Nevertheless, the PED diagram also shows intense diffraction spots each
60$^{\circ}$ in $\phi$ at $\theta =$ 60$^{\circ}$ [white ellipses on Fig.
\ref{fig1}b)], as well as dark hexagonal-shaped diffraction lines [dark, dark
grey and grey hexagons on Fig. \ref{fig1}b)] of six-fold symmetry.
Interestingly, both the bright diffraction spots and the dark diffraction lines
are well reproduced in our calculations for the BLG with interlayer spacing
3.46~\AA\ \footnote{Note that we did not attempt to optimize the calculation
parameters for improving the experiment-theory agreement since that work was
already well done in Ref. \cite{Razado2018}.} regardless of whether Bernal
[Fig. \ref{fig1}c)] or AA-stacking [Fig. \ref{fig1}d)] is assumed. These PED
features are fingerprints of the honeycomb lattice of an isolated monolayer
graphene and originate from purely in-plane scattering. Indeed, the very low
atomic number of carbon, its small scattering power and the large interlayer
separation characteristic of layered materials implies that in-plane coherent
scattering is much larger than out-of-plane scattering \cite{Kuttel1994,
Maillard1996}. In particular, the sharp hexagonal-shaped dark diffraction lines
are reminiscent of destructively overlapping circular diffraction rings
\cite{Greber1998} already observed in graphite as well as boron nitride
\cite{Auwarter1999, Matsui2012, Roth2013}, and appearing around the
forward-focusing peaks associated with in-plane single scattering by C atoms of
successive coordination shells around the emitter. The opening angles of the
associated interference cones depend on the emitter-scatterer distance, the
photoelectron wavenumber, and the scattering phase shift which is particularly
significant for an $s$-wave on carbon at this kinetic energy \cite{Matsui2012}.
A complete description of these diffraction rings is beyond the scope of the
present paper and has been already done elsewhere. Nevertheless, our
simulations performed for a graphene layer have confirmed that they can be well
reproduced at the single-scattering level considering eight coordination shells
(that corresponds to an emitter scatterer distance of 5.12~\AA) around the
carbon emitter (results not shown). For the purpose of the present paper, we
keep in mind that these peculiar PED features are all associated with
interference of the direct electronic wave with those single-scattered along
the dense C-C atomic directions of the six-fold symmetric graphene lattice.

We now focus on PED features that carry information about the stacking
geometry. Specifically we look at the signal from electrons photoemitted from a
bottom layer C atom and scattered by a top layer atom. We see that the
simulated PED stereographic projection of the Bernal-stacked BLG exhibits a
clear three-fold symmetry [Fig. \ref{fig1}c)], whereas the one of the AA
stacking [Fig. \ref{fig1}d)] shows bright sixfold-symmetric diffraction spots
at $\theta =$ 32.5$^{\circ}$ that do not appear in experiment. This is fully
consistent with the SE-odd, respectively SE-even parities of the Bernal and AA
stacking of BLG whose unit cells have either half [Fig. \ref{fig1}e)], or all
[Fig. \ref{fig1}f)] of the sublattices sites eclipsed. It turns out that, even
if in-plane single scattering clearly dominates at this kinetic energy, the BLG
stacking sequence can clearly be distinguished by looking at the PED pattern.
The only experimental features that are not captured in the simulated PED
stereographic projection of Bernal-stacked BLG are the six-fold symmetric
bright spots at $\theta=17.6^{\circ}$. They are most likely due to the trilayer
stacking made of Bernal BLG and the carbon buffer layer when grown on the
SiC(0001) substrate. We conclude from this section that PED of BLG with a
kinetic energy as high as $\sim$640~eV, can be reproduced accurately and with a
high sensitivity to the stacking geometry with the present full-multiple
scattering simulations.   

\subsection{Layer-resolved PED of AA-stacked BLG.}

\begin{figure}[ht!]
\includegraphics[width=0.45\textwidth]{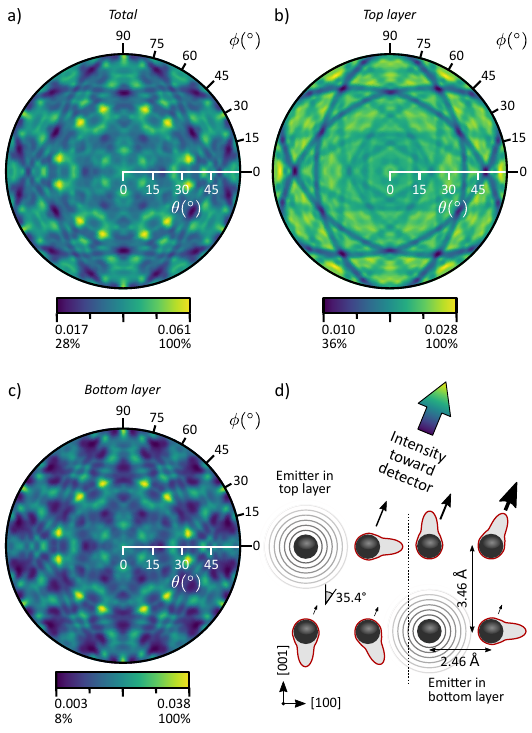}
\caption{
    [color online] a) Single-scattering calculation of the stereographic
    projection of PED from C 1$s$ at 638.6 eV kinetic energy for AA-stacked BLG.
    b),c) Corresponding stereographic projections of PED for the top and bottom
    layers of the stacking, respectively. d) Sketch of first nearest neighbors
    atoms in the \{$\bar{1}20$\} planes showing dominant scattering paths, as
    indicated by the size of the black arrows, for inter- and intra-layer
    scattering. The scattering factors calculated at 638.6 eV kinetic energy are
    plotted in red. The detector direction is fixed along the interlayer forward
    focusing direction at $\theta =$35.4$^{\circ}$. }
    \label{fig2}
\end{figure}

We now focus on the separated contributions of each layer to the total PED
pattern of AA-stacked BLG. Layer resolution is a trivial problem from a
theoretical point of view, since in core-level PED the photoemission
intensities of different emitter atoms add incoherently. In experiment, layer
resolution can be achieved by numerical deconvolution of core levels exhibiting
well-defined chemical shifts. It has been reported for BLG grown on SiC(0001)
\cite{Riedl2010, Razado2018}, or transferred on SiO$_2$ \cite{Kim2008,
deLima2013, Matsui2013}. Figure \ref{fig2}a) shows the PED stereographic
projection of AA-stacked BLG calculated at the single-scattering level. Single
scattering calculations greatly reduce computation time while preserving the
main features of the full mulitple scattering result [Fig. \ref{fig1}d)].
Indeed, both the hexagonal-shaped dark diffraction lines characteristic of
single-layer graphene at large polar angles and the distinguishing features of
the AA-stacking at smaller angles are well reproduced in the single-scattering
calculation [Fig. \ref{fig2}a)]. In particular, the bright sixfold-symmetric
diffraction spots at $\theta =$ 32.5$^{\circ}$ can be interpreted as belonging
to the sides of typical volcano-shaped PED structures centered around the
interlayer forward focusing direction at $\theta =$35.4$^{\circ}$. As explained
for other systems, this volcano shape is strongly energy-dependent and emerges
from interference phenomena due to atoms close to dense forward scattering
directions \cite{Agliz1995, Juillaguet1995, Schieffer2001, Tricot2022}.   

The layer-resolved PED stereographic projections of the top [Fig. \ref{fig2}b)]
and bottom [Fig. \ref{fig2}c)] layers of the AA-stacked BLG, now allow us to
show that the full information about the stacking geometry is brought by PED
from the bottom layer only, the PED pattern of the top layer being almost
identical to the one of an isolated single layer graphene (result not shown).
In contrast to the PED pattern of the bottom layer which is dominated by
inter-layer forward scattering events, the PED pattern of the top layer is
almost purely constructed from interfering in-plane single scattering events.
Indeed, considering an emitter atom in the top layer, the scattering paths
carrying information about the stacking geometry of BLG have to imply
out-of-plane backscattering on the bottom layer for allowing the photoelectrons
to be detected. So, even though such scattering paths do exist, one can
understand, by looking at the scattering factor which is rather forward-peaked
at this kinetic energy, that these back-scattering events have negligible
intensity compared to the intra-layer scattering [Fig. \ref{fig2}d), left-hand
side]. In contrast, the probability for an electron photoemitted from the
bottom layer to suffer only one scattering process on a top-layer atom before
reaching the detector is much larger than the probability for this electron to
undergo an additional in-plane scattering event [see the Fig. \ref{fig2}d),
right-hand side].    

Our layer-resolved study thus highlights drastically different PED patterns for
the bottom and top layers of BLG, the former providing information on the SE
parity of the stacking and the latter being similar to single-layer graphene.
As we will now see in the next sections by introducing twist angles, this not
only offers an appealing way for accurately determining the relative
orientations of two stacked graphene layers, but also for resolving the
internal atomic arrangement of moiré patterns on the entire possible range of
twisted angles, from small magic angles up to the $30^{\circ}$-twisted
quasicrystal.

\subsection{Twisted BLG.}

We start by demonstrating that introducing rotational misalignment between the
two layers of BLG, while working in the forward-scattering regime of PED,
completely modifies the PED pattern of the bottom layer and allows a precise
determination of the twist angle. First of all, we constructed a twisted BLG
structure at $21.79^{\circ}$ commensurate angle corresponding to translation
indices $(p,q)=(1,2)$ and giving rise to the smallest commensurate hexagonal
Moiré supercell with a 6.51~\AA\ basis vector length and containing 28 atoms
[see Table \ref{tbl:pq}]. To do so, we considered the perfectly eclipsed
honeycomb lattices of the AA-stacked BLG and rotated counter-clockwise the top
layer around coinciding carbon atoms of the two graphene sublattices.

\begin{figure}[ht!]
\includegraphics[width=0.45\textwidth]{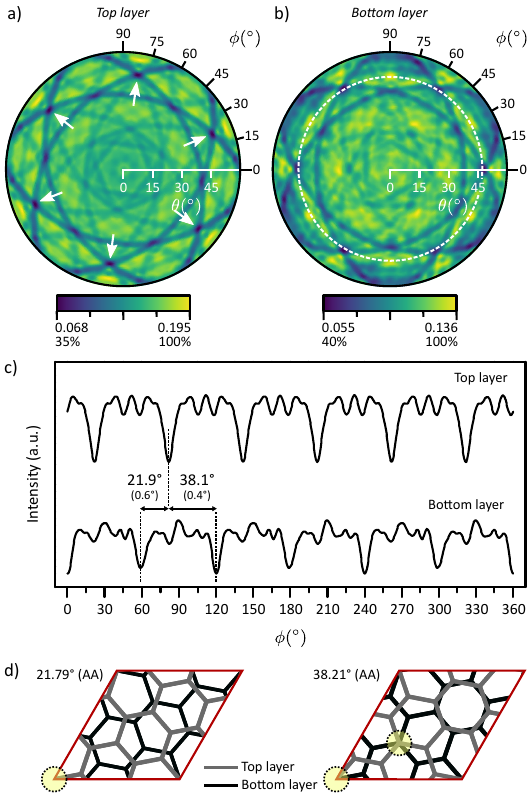}
\caption{
    [color online] a,b) Layer-resolved single scattering-calculated
    stereographic projections of PED from C 1$s$ at 638.6~eV kinetic energy for
    a twisted BLG at 21.79\degree\ commensurate angle. c) Azimuthal profiles
    for top and bottom layers at 47.5\degree\ polar angle [see the white-dashed
    circle in b)]. The white arrows in a) indicate the PED intensity
    depressions appearing at the crossing points of the hexagonal black
    diffraction lines. They are manifested as six sharp inverted peaks in the
    azimuthal PED profiles in c) and are used as reference for the twist angle
    determination (standard deviation in parentheses). d) Twisted BLG
    supercells of 21.78\degree\ and 38.21\degree\ commensuration partners.
    Whereas the unit cells are of equal areas they exhibit distinct SE parities
    as highlighted by the yellow circles that shows the coincident sublattice
    sites. One sublattice site is eclipsed in SE-odd commensuration cell
    (21.78\degree, starting from AA stacking). Both A and B sublattice sites of
    each layer coincide with atomic sites of the neigbouring layer in SE-even
    commensuration cell (38.21\degree, starting from AA stacking).}
\label{fig3}
\end{figure}

Figures \ref{fig3}a)-b) show the layer-resolved single scattering-calculated
stereographic projections of PED from C 1$s$ at 638.6~eV kinetic energy for the
twisted BLG. While the PED of the top layer [Fig. \ref{fig3}a)] is identical to
the PED stereographic projection of the top layer of the AA-stacked BLG [Fig.
\ref{fig2}b)], i.e. of single layer graphene (apart a rotation corresponding to
the introduced twist angle since referenced to the bottom layer), we
immediately see that the bottom layer of twisted BLG now exhibits a drastically
different PED pattern from its AA-stacked counterpart [Fig. \ref{fig2}c)]. All
the PED features characteristic of interlayer forward scattering have
disappeared and the PED pattern is similar to that of the top layer of twisted
BLG [Fig. \ref{fig3}a)], except for more blurred diffraction lines and a global
rotational misalignment. 

In fact, due the multiplicity of atomic environments and associated upward
scattering directions around the emitting atom resulting from the twist, the
total PED intensity signal that was before concentrated around well-defined
dense forward-scattering directions becomes redistributed among many polar and
azimuthal angles leading to a strongly reduced PED anisotropy over the full
stereographic projection. As a result, the in-plane single forward-scattering
events that hardly contribute to the total PED intensity in a simple stacking
geometry, become important and dictate the PED anisotropy in twisted BLG.
Interestingly, this crossover from dominant inter- to intra-layer forward
scattering regime concomitantly offers an appealing experimental way to
determine misorientation angles of twisted BLG in high-energy PED since the PED
patterns of the top and bottom layers are both characteristic of an isolated
single layer graphene and differ essentially only by a relative rotation.
Figure \ref{fig3}c) shows azimuthal PED profiles taken for both the top and
bottom layers at fixed polar angle value of 47.5\degree\ [white-dashed circle
in Fig. \ref{fig3}b)]. This polar angle was chosen because it allows to focus
on the PED intensity depressions appearing at the crossing points of the
hexagonal black diffraction lines [see white arrows in Fig. \ref{fig3}a)]. They
show up as six sharp inverted peaks in the azimuthal PED profiles [Fig.
\ref{fig3}c)], providing an ideal spectroscopic fingerprint to access the twist
angle value with a reasonable precision. Indeed, the inverted peaks of the
azimuthal PED profile of one given layer always appear at
21.9$^{\circ}$$\pm$0.6$^{\circ}$ and 38.1$^{\circ}$$\pm$0.4$^{\circ}$ from two
consecutive ones of the other layer. These two angle values in fact correspond
to the commensuration partners $\theta_{t}$ and $60^{\circ}-\theta_{t}$ that
lead to unit cells of equal areas but exhibit distinct SE parities
\cite{Mele2010, Mele2012} [see Fig. \ref{fig3}d)]. Interestingly, the
three-fold symmetry of our computed SE-odd BLG twisted stacking with
translation indices $(p,q)=(1,2)$ is also tracked in the PED of the bottom
layer. Whereas the top layer exhibits a PED stereographic projection identical
to the one of single layer graphene and inverted peaks of equal amplitude in
the azimuthal profile, the PED stereographic projection of the bottom layer
shows a slight 60\degree-anisotropy of the intensity at small polar angles
($\theta\le30^{\circ}$) [Fig. \ref{fig3}b)], and the azimuthal profile exhibits
inverted peaks of equal amplitude every 120\degree, only [bottom profile, Fig.
\ref{fig3}c)]. 

\begin{table}[b]
\centering
\begin{tabular}{p{.1\textwidth}p{.05\textwidth}p{.05\textwidth}p{.05\textwidth}r}
	\hline
	{$\theta_{t}$ (\degree)} & $p$ & $q$ & $N$ & {$||\overrightarrow{t'_i}||$ (\AA)} \\
	\hline
	\hline
    0.00  & 1 & 1 & 4 & 2.46 \\	
    1.08  & 30 & 31 & 11164 & 129.96 \\
    3.89  & 8 & 9 & 868 & 36.24 \\
    4.41  & 7 & 8 & 676 & 31.98 \\
	13.17 & 2 & 3 & 76  & 10.72 \\
    16.43 & 3 & 5 & 196 & 17.22 \\
    17.90 & 4 & 7 & 124 & 13.70 \\
	21.79 & 1 & 2 & 28  & 6.51 \\
	25.04 & 4 & 9 & 532 & 28.37 \\
    26.01 & 3 & 7 & 316 & 31.98 \\
    27.80 & 2 & 5 & 52  & 8.87 \\
	\hline
\end{tabular}
\caption{
    Selected commensurate angles $\theta_{t}$ in a 30$^{\circ}$-wide
    interval of twist angles. $(p,q)$, $N$, and $||\overrightarrow{t'_i}||$ are the
    translation indices, number of atoms and length of the primitive translation
    $\overrightarrow{t'_i}$ of the associated commensuration supercell.}
\label{tbl:pq}
\end{table}

\begin{figure*}[t]
\includegraphics[width=1.0\textwidth]{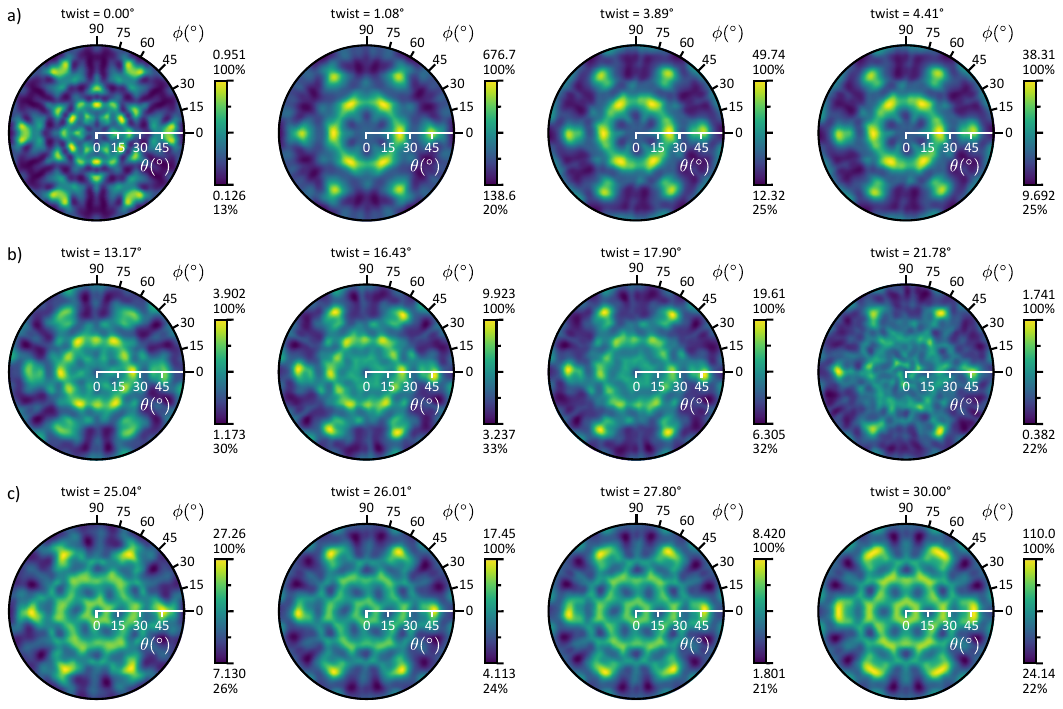}
\caption{
    [color online] Bottom layer stereographic projections of PED from C 1$s$ at
    100~eV kinetic energy of twisted BLG calculated in full multiple scattering.
    Three distinct regimes are observed depending on the twist angles: a) small
    angles regime from 0\degree\ (AA stacking) to 4.41\degree, b) Moiré crystal
    regime from 13.17\degree\ to 21.78\degree\ and c) quasicrystal regime from
    25.04\degree\ to 30.00\degree.}
\label{fig4}
\end{figure*}

Having demonstrated that layer-resolved high-energy PED is a particularly
well-suited experimental technique for accessing both twist angle values and SE
parities of moiré supercells in BLG, let us now go a step further and see if
PED also yields structural information about the internal atomic arrangements
of the moiré patterns themselves. To this end, we now focus on low-energy PED
by working at 100 eV kinetic energy within the full-multiple scattering
approach. Indeed, at such low kinetic energy, the scattering factor is more
isotropic and the single-scattering approximation is no longer sufficient.
Together with the extremely short inelastic mean free path of photoelectrons
(5.8~\AA) which prevents the formation of well-defined diffraction fringes, one
enters in an ideal PED regime for obtaining structural information about the
atomic configurations within the moiré supercells. Furthermore, working at such
a low kinetic energy allows us to explore a large number of calculation
parameters and atoms in the cluster with reasonable computation times and
limited memory resources ($\leq$100 GB), which is crucial for large moiré
supercells. Figure \ref{fig4} shows the full-multiple scattering calculated PED
stereographic projections of the BLG bottom layer from C 1$s$ at 100~eV kinetic
energy for various moiré supercells within a twist angles range of 30$^{\circ}$
[see Table \ref{tbl:pq}] \footnote{Note that we will not discuss the top layer
of BLG anymore. Indeed, at low kinetic energy, even though the top layer PED
pattern is constructed from multiple scattering events that are more isotropic
in space, non stacking-sensitive in-plane multiple scattering remains much more
dominant.}.  We are mostly interested in twist angles that lead to small
commensurate supercells, but for the sake of completeness with regard to the
rich physics of twisted BLG, we consider two more strutures: (i) one with
$\theta_{t}$=1.08\degree, corresponding to the magic angle \cite{Cao2018a},
which gives rise to a supercell containing as many as 11164 atoms [Fig.
\ref{fig4}a), second panel], and (ii) the purely incommensurate dodecagonal
quasi-crystal with $\theta_{t}$=30\degree\ [Fig. \ref{fig4}c), last panel]. As
before, at large polar angles ($\theta\ge30^{\circ}$), all PED patterns share
many similar features, which may be attributed to in-plane multiple scattering
events and which are inherited from single layer graphene (result not shown).
At small polar angles ($\theta\le30^{\circ}$), however, one can indentify three
types of closely resembling PED patterns associated with small [Fig.
\ref{fig4}a)], intermediate [Fig. \ref{fig4}b)] and large twisted angles [Fig.
\ref{fig4}c)]. At small twist angles, all PED patterns exhibit a six-fold
symmetry and rather broad bright spots lying on diffuse ring of intensity each
60\degree\ in $\phi$ at $\theta=22.3\degree$. In contrast, the pattern for
intermediate twist angles have six- or three-fold symmetry [see e.g. Fig.
\ref{fig4}b), last panel] and show PED spots which are more structured and
resolved from the diffuse intensity ring as reminiscent of the sharp PED
features of the AA-stacked BLG [Fig. \ref{fig4}a), first panel]. Last, for the
largest twist angles, a rather diffuse and robust honeycomb-like PED pattern is
always found [Fig. \ref{fig4}c)].  

\begin{figure*}[ht!]
\includegraphics[width=1\textwidth]{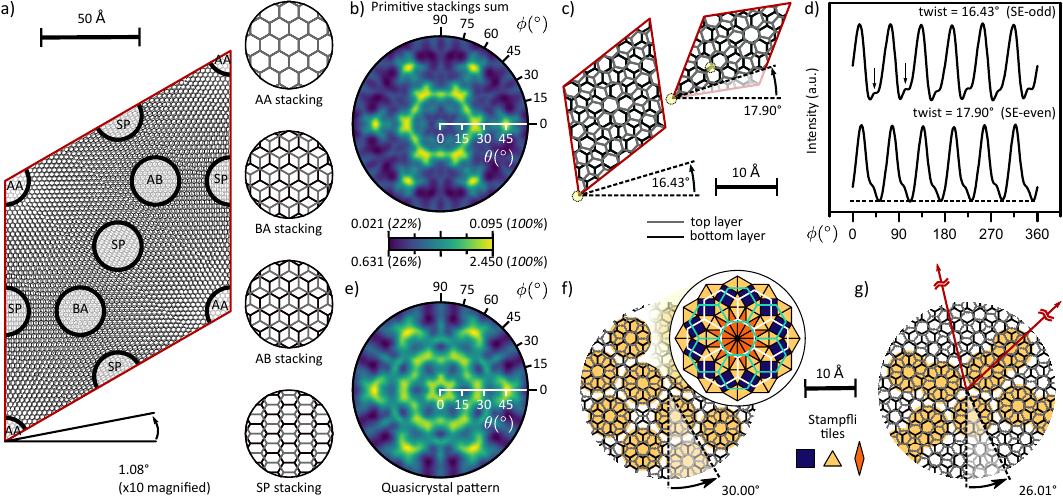}
\caption{
    [color online] a) Moiré supercell of BLG (red parallelogram) generated by
    the magic-twist angle ($p=30$, $q=31$). The atomic registry within such a
    large-period superlattice ($\sim$13 nm) can be viewed as evolving smoothly
    between widely separated regions with locally AB, BA, AA and SP stackings. b)
    Full-multiple scattering calculated PED stereographic projection from C 1$s$ at
    100~eV kinetic energy corresponding to the incoherent sum of PED patterns
    associated with AA, AB, BA and SP stackings. c) Twisted BLG supercells of
    SE-odd and SE-even commensurate moiré crystals at 16.43\degree\ ($p=3$, $q=5$)
    and 17.90\degree\ ($p=4$, $q=7$) of twist angle. The yellow areas highlight
    coincident sublattice sites. Whereas only one sublattice site is eclipsed in
    SE-odd commensuration cell, both A and B sublattice sites of each layer
    coincide with atomic sites of the neigbouring layer in SE-even commensuration
    cell. d) Integrated azimuthal profiles from 0\degree\ to 15\degree\ in polar
    angle for 16.43\degree\ (top) and 17.90\degree\ (bottom) twisted BLG evidencing
    the sensitivity to the SE parity. e) Full-multiple scattering calculated PED
    stereographic projection from C 1$s$ at 100~eV kinetic energy of the
    characteristic tiling located at the center of rotation in the Stampfli
    tessellation of the 12-fold quasicrystal. f-g) Twisted BLG superlattices of the
    dodecagonal quasicrystal and of a large commensurate moiré crystal at
    30.00\degree\ and 26.01\degree\ ($p=3$, $q=7$) of twist angle, respectively.
    The orange areas in f) and g) highlight the 48-atoms subset of the Stampfli
    tessellation considered in the calculation of e).}
\label{fig5}
\end{figure*}

Let us now discuss these three identified PED regimes in terms of internal
atomic arrangements of moiré superlattices. Figure \ref{fig5}a) shows the moiré
supercell of BLG (red parallelogram) generated by the magic-twist angle
($p=30$, $q=31$). The atomic registry within such a large-period superlattice
($\sim$13 nm) can be viewed as evolving smoothly between widely separated
regions with locally AB, BA, AA and SP (saddle point) stackings. This is
typical of superlattices described by a small twist. The larger the twist
angle, the smaller the superlattice period and total area proportion of atoms
belonging to a well-defined stacking. Figure \ref{fig5}b) shows a full-multiple
scattering calculated PED stereographic projection corresponding to the
incoherent sum of PED patterns associated with AA, AB, BA and SP stackings,
only. Interestingly, it closely matches the PED patterns obtained for small
twist angles [Fig. \ref{fig4}a)] by considering all the inequivalent emitter
atoms of the supercells. Both the six-fold symmetry and bright spots lying on
the diffuse ring of intensity at $\theta=22.3\degree$ are well reproduced.
While the former can be understood as emerging from the mixed six-fold,
three-fold with mirror plane and two-fold symmetries of 120\degree\ equivalent
domains of the AA, AB/BA and SP stackings respectively, the latter corresponds
to the out-of-plane nearest neighbors forward scattering atomic direction
existing in all primitive BLG stackings. This demonstrates, that, for such a
range of twist angles, the whole structural information about the atomic
arrangement within the moiré superlattice is contained within small clusters of
atoms belonging to stackings very close to those of primitive BLG. While this
offers appealing computational means for drastically reducing the memory
ressources when performing full-multiple scattering simulations of small twist
angles superlattices, it also demonstrates that low-energy PED will not show a
high sensitivity to the twist angle as long as these structures do not
correspond to commensurate moiré crystals of specific atomic configurations and
SE parities. Figure \ref{fig5}c) shows twisted BLG supercells of SE-odd and
SE-even commensurate moiré crystals at twist angles 16.43\degree\ and
17.90\degree, respectively. For such close $\theta_{t}$ values associated with
rather large supercells, we can see that, although the PED patterns look very
similar [Fig. \ref{fig4}b)], they do carry information about the SE parity.
Indeed, the extracted azimuthal profiles integrated from 0\degree~to
15\degree~in polar angle exhibit slight shoulders on the left-hand side of the
main PED peaks of either 120\degree\ [top azimuthal profile, Fig. \ref{fig5}d)]
or 60\degree\ [bottom azimuthal profile, Fig. \ref{fig5}d)] periodicity as
expected for the odd and even SE parities of the commensurate moiré crystals
associated with the translation indices ($p$, $q$)=  (3, 5) and (4, 7),
respectively. 

Note that the sensitivity to the SE parity in this intermediate range of twist
angles strongly depends on the size of the commensurate moiré supercell. As we
can see on Fig. \ref{fig4}b), the three-fold symmetry of the smallest SE-odd
moiré supercell for a twist angle of 21.78\degree, is easily tracked over the
first 30\degree\ of the PED stereographic projection without the need for a
highly-resolved azimuthal profile. In this commensurate moiré supercell, the
emitter atom that carries the SE-odd parity counts for 1/14 of the full PED
signal which, apart for the Bernal-stacked BLG, is the highest achievable
intensity ratio. 

\begin{figure}[ht!]
\includegraphics[width=0.45\textwidth]{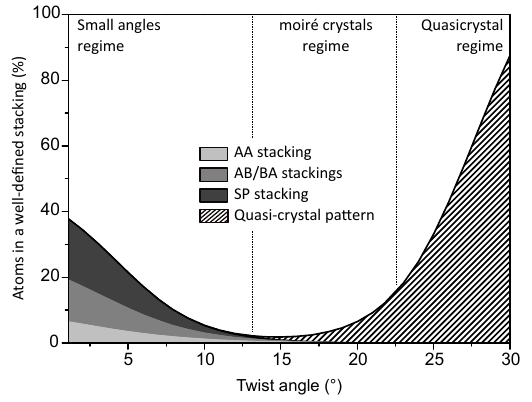}
\caption{
    PED phase diagram of twisted BLG: proportion of atoms belonging in a
    well-defined stackings of either primitive BLG [Fig. \ref{fig5}a)] or Stampli
    tessellation of BLG quasi-crystal [Fig. \ref{fig5}f)] as a function of
    commensurate twist angles from 1\degree\ to 30\degree. The boundaries of the
    various regimes well correspond to those identified in Fig.~\ref{fig4}.}
\label{fig6}
\end{figure}

Turning now to larger twist angles, we saw that we quickly entered in the PED
regime dictated by a robust honeycomb-like PED pattern. Similar to the stacking
analysis that we performed for the small twist angles, we can show that such a
peculiar pattern arises from the specific atomic configuration of the 12-fold
Stampfli tiling of the BLG quasicrystal. Figure \ref{fig5}e) shows a PED
stereographic projection calculated considering only a 48-atoms subset of the
characteristic tiling located at the center of rotation in the Stampfli
tessellation [see Fig.\ref{fig5}f)] \cite{Sadoc2023}. It almost perfectly
matches not only the full-multiple scattering calculation of the
30\degree-twisted BLG [last panel, Fig. \ref{fig4}c)] but also the PED pattern
calculated for the whole considered range of large twist angles. This
demonstrates that the honeycomb-like PED pattern directly reflects the ubiquity
of atomic arrangements closely matching the one of the 12-fold Stampfli tiling
as seen in Fig.\ref{fig5}g). It is also worth mentioning that even if the
quasi-crystal lacks translational symmetry, its rotational symmetry as well as
its fractal nature allows to restrict PED calculations to small clusters of
atoms containing the elementary building block of the Stamplfi tessellation and
of diameter roughly corresponding to the inelastic mean free path. We close
this section by summarizing our results in a \textit{PED phase diagram of
twisted BLG} [Fig. \ref{fig6}]. It is based on the proportion of atoms
belonging to a well-defined stacking of either primitive BLG or Stampli
tessellation of BLG quasi-crystal as a function of the twist angle. To do so,
we first consider each emitter as well as its three first neighbors in the same
layer. We then consider a local AA stacking if, for these four atoms, one lies
above another one of the other layer within a tolerance radius corresponding to
half the atomic radius of carbon ($r_t = 0.35$~\AA). For accounting for the
AB/BA and SP stackings, one simply applies the translation needed to recover an
AA stacking before checking the interlayer atomic coincidence. By applying this
procedure, an atom is either identified as AA, AB/BA, SP or none of the three,
but never falls into more than one of these categories. For the stacking
characteristic of the quasi-crystal, we consider 24 atoms of the pattern
identified in figure~\ref{fig5}f) centered on the hexagon to which the emitter
atom belongs to and rotate them by 30\degree\ before testing if all of these 24
atoms can match their in-plane position with 24 atoms of the second layer using
the same tolerance criterion as above. Figure \ref{fig6} shows the so-obtained
PED phase diagram of twisted BLG. It validates the tolerance criterion we
applied since it very well tracks the three PED regimes that we have identified
above: a \textit{magic-angle like regime} at small twist angles
($0^{\circ}\le\theta_{t}\le13^{\circ}$) where PED tracks the proportion areas
of relatively well-defined AA, AB, BA and SP stackings, \textit{a moiré crystal
regime} at intermediate twist angles ($13^{\circ}\le\theta_{t}\le23^{\circ}$)
in which the SE-parities of the commensurate moiré supercells are probed and a
\textit{moiré quasi-crystal regime} at large twist angles
($23^{\circ}\le\theta_{t}\leq30^{\circ}$) that reflects the characteristic
tiling of the 12-fold Stampfli tessellation.

\subsection{Sensitivity of modulation functions to interlayer distances of primitive and twisted BLG.}

\begin{figure}[t]
\includegraphics[width=0.45\textwidth]{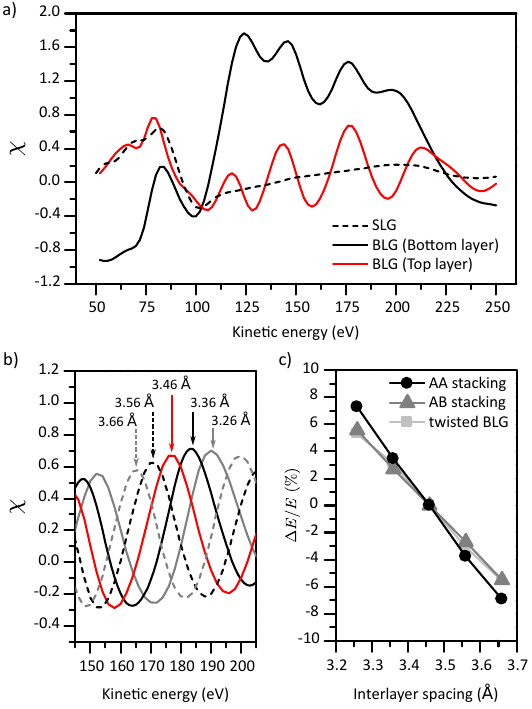}
\caption{
    [color online] a) C $1s$ normal-emission modulation function as a function
    of kinetic energy of the top (red curve, $\chi_{top}(E)$) and bottom (black
    curve, $\chi_{bottom}(E)$) layers of AA-stacked and of single layer graphene
    (black dashed curve). b) Zoom-in of $\chi_{top}(E)$ around the modulation peak
    at $\sim$175 eV for various interlayer distance in a range of $\pm$0.2~\AA\
    around 3.46~\AA. c) Relative energy shifts of modulation maxima of
    $\chi_{top}(E)$ as a function of the interlayer distances for AA (black
    circles), Bernal (dark grey triangles), and 21.8\degree-twisted BLG (light grey
    squares).}
\label{fig7}
\end{figure}

In this last section, we illustrate how sensitive the energy-dependence of the
modulation function of the C 1$s$ cross-section is to tiny changes of the
interlayer distance in both primitive and twisted BLG. We consider the
modulation function at normal-emission ($\theta=0\degree$), calculated in
full-multiple scattering for kinetic energies of photoelectrons ranging from 50
to 250 eV, i.e., in the low-energy backscattering regime of PED
\cite{Woodruff1992, Woodruff1994}. The modulation function $\chi (E)$ is
defined by \cite{Tricot2022}:

\begin{eqnarray}
{\chi (E)=\frac{I_0 (E)-I_{0, direct} (E)}{I_{0, direct} (E)}}, \label{eq1}
\end{eqnarray}

where $I_0 (E)$ and $I_{0, direct}$ are respectively the calculated, total and
direct (diffraction-free), normal photoemission intensities at a given kinetic
energy $E$. Figure \ref{fig7}a) shows the modulation functions of the top (red
curve, $\chi_{top}(E)$) and bottom (black curve, $\chi_{bottom}(E)$) layers of
AA-stacked BLG as well as the one of single layer graphene (black-dashed curve)
for comparison. All the modulation functions show an oscillating behavior that
arises from coherent interference phenomena of different scattering paths, the
relative phases changing due to the varying electron wavelength, and positive
(negative) values of $\chi(E)$ corresponding to purely constructive
(destructive) interference. For single layer graphene, where all interference
is due to in-plane scattering, $\chi(E)$ exhibits a long-period oscillation of
$\sim$130 eV corresponding to the graphene lattice parameter of 1.42~\AA\ in
real-space. Focusing on BLG, we first see that the in-plane nearest-neighbors
forward scattering events that give rise to the long-period oscillation still
contribute to $\chi_{bottom}(E)$ and $\chi_{top}(E)$ as a broad background.
Yet, they both show additional modulations of smaller periods reflecting
out-of-plane scattering events. While $\chi_{bottom}(E)$ results from the
combination of one main modulation of frequency $\sim$60 eV with a secondary
one of half period, $\chi_{top}(E)$ is made of one single frequency of $\sim$30
eV superimposed to the in-plane long-period oscillation. When these kinetic
energy oscillation frequencies are transposed to real-space, they correspond to
the interlayer distance (3.46~\AA) for the 60 eV period and twice the
interlayer distance (7.92~\AA) for 30 eV. In fact, for the top layer, the
normal-emission configuration only selects the 180\degree~backscattering
geometry resulting in a modulation function dominated by a single oscillation
of frequency corresponding to twice the value of the inter-layer distance
\cite{Woodruff1994}. For the bottom layer, it also takes into account the
favorable out-of-plane forward scattering carrying the path difference
associated to one vdW gap. Figure \ref{fig7}b) shows a zoom-in of
$\chi_{top}(E)$ around the modulation peak at $\sim$175 eV for various
interlayer distance of the AA-stacked BLG in a range of $\pm$0.2~\AA\ around
3.46~\AA. We see that the modulation peak experiences rather large negative or
positive energy shifts for increased or decreased interlayer distances,
respectively. The relative energy shift $\Delta E/E$ of the modulation peak
depends almost linearly on the interlayer spacing [dark curve with circles on
Fig. \ref{fig7}c)], with a slope of about 40~\%/\AA\ demonstrating the
remarkable sensitivity of backscattering energy-scans to small changes of the
interlayer distances. This applies not only to primitive BLG, where the
modulation function of the top layer of the Bernal-stacking shows identical
dependence on the interlayer distance [heavy-grey curve with triangles on Fig.
\ref{fig7}c)], but also to twisted BLG as demonstrated for the
21.79\degree-twisted commensurate BLG supercell [light-grey curve with squares
on Fig. \ref{fig7}c)].

\section{Conclusion}
 
By theoretically analyzing the PED pattern formation of twisted BLG with
single- and multiple scattering calculations, we have thoroughly examined the
capacity of PED for accessing, at the sub-\AA\ length scale, local structural
properties of moiré superlattices made of twisted vdW materials. We have shown
that high-energy PED can be used for determining both twist angle values and SE
parities of moiré supercells, whereas low-energy PED allows to obtain
structural information about the internal atomic arrangements of the moiré
superstructures and to probe interlayer distances in energy-scan mode. We hope
that these theoretical findings will motivate experimentalists to perform PED
with high angle, energy and spatial resolution for deeply probing the atomic
structure of the vast and fascinating family of twisted, strained, and
intercalated vdW materials.

\section*{CRediT authorship contribution statement}
\textbf{S. Tricot}: Writing-review editing, Methodology, Investigation, Formal
analysis, Supervision, Resources, Data curation, Software. \textbf{H. IKeda}:
Investigation, Review editing. \textbf{H.-C. Tchouekem}: Investigation, Review
editing. \textbf{J.-C. Le Breton}: Investigation, Review editing. \textbf{S.
Yasuda}: Investigation, Review editing. \textbf{P. Krüger}: Investigation,
Writing-review editing, Supervision. \textbf{P.  Le F\`evre}:  Investigation,
Writing-review editing. \textbf{D. S\'ebilleau}: Project administration,
Software, Review editing. \textbf{T. Jaouen}: Writing – review editing, Writing
– original draft, Methodology, Investigation, Formal analysis, Supervision.
\textbf{P. Schieffer}: Project administration, Supervision, Methodology,
Resources, Investigation, Formal analysis. 

\section*{Declaration of Competing Interest} 
The authors declare that there are no known conflicts of interest and no
significant financial support associated with this work that could have
influenced its outcome.  

\section*{Data availability}
Data will be made available on request.

\section*{Acknowledgments}
S. Tricot, J.-C. Le Breton, D. S\'ebilleau, T. Jaouen and P. Schieffer
acknowledge partial funding from Horizon Europe MSCA Doctoral network grant
n.101073486, EU-SpecLab, funded by the European Union. H.-C. Tchouekem and T.
Jaouen acknowledge the support of the French National Research Agency (ANR)
(MOSAICS project, ANR-22-CE30-0008). All the authors thank J.  Gardais and G.
Raffy for their technical support on the computing cluster. 


\begin{thebibliography}{10}
\expandafter\ifx\csname url\endcsname\relax
  \def\url#1{\texttt{#1}}\fi
\expandafter\ifx\csname urlprefix\endcsname\relax\def\urlprefix{URL }\fi
\expandafter\ifx\csname href\endcsname\relax
  \def\href#1#2{#2} \def\path#1{#1}\fi

\bibitem{Kim2016}
K.~Kim, M.~Yankowitz, B.~Fallahazad, S.~Kang, H.~C.~P. Movva, S.~Huang,
  S.~Larentis, C.~M. Corbet, T.~Taniguchi, K.~Watanabe, S.~K. Banerjee, B.~J.
  LeRoy, E.~Tutuc, \href{https://doi.org/10.1021/acs.nanolett.5b05263}{van der
  {Waals} {Heterostructures} with {High} {Accuracy} {Rotational} {Alignment}},
  Nano Letters 16~(3) (2016) 1989--1995.

\bibitem{Ribeiro2018}
R.~Ribeiro-Palau, C.~Zhang, K.~Watanabe, T.~Taniguchi, J.~Hone, C.~R. Dean,
  \href{https://www.science.org/doi/full/10.1126/science.aat6981}{Twistable
  electronics with dynamically rotatable heterostructures}, Science 361~(6403)
  (2018) 690--693.

\bibitem{Cao2018a}
Y.~Cao, V.~Fatemi, S.~Fang, K.~Watanabe, T.~Taniguchi, E.~Kaxiras,
  P.~Jarillo-Herrero,
  \href{https://www.nature.com/articles/nature26160}{Unconventional
  superconductivity in magic-angle graphene superlattices}, Nature 556~(7699)
  (2018) 43--50.

\bibitem{Cao2018b}
Y.~Cao, V.~Fatemi, A.~Demir, S.~Fang, S.~L. Tomarken, J.~Y. Luo, J.~D.
  Sanchez-Yamagishi, K.~Watanabe, T.~Taniguchi, E.~Kaxiras, R.~C. Ashoori,
  P.~Jarillo-Herrero,
  \href{https://www.nature.com/articles/nature26154}{Correlated insulator
  behaviour at half-filling in magic-angle graphene superlattices}, Nature
  556~(7699) (2018) 80--84.

\bibitem{Yankowitz2019}
M.~Yankowitz, S.~Chen, H.~Polshyn, Y.~Zhang, K.~Watanabe, T.~Taniguchi,
  D.~Graf, A.~F. Young, C.~R. Dean,
  \href{https://www.science.org/doi/full/10.1126/science.aav1910}{Tuning
  superconductivity in twisted bilayer graphene}, Science 363~(6431) (2019)
  1059--1064.

\bibitem{Sharpe2019}
A.~L. Sharpe, E.~J. Fox, A.~W. Barnard, J.~Finney, K.~Watanabe, T.~Taniguchi,
  M.~A. Kastner, D.~Goldhaber-Gordon,
  \href{https://www.science.org/doi/full/10.1126/science.aaw3780}{Emergent
  ferromagnetism near three-quarters filling in twisted bilayer graphene},
  Science 365~(6453) (2019) 605--608.

\bibitem{Lu2019}
X.~Lu, P.~Stepanov, W.~Yang, M.~Xie, M.~A. Aamir, I.~Das, C.~Urgell,
  K.~Watanabe, T.~Taniguchi, G.~Zhang, A.~Bachtold, A.~H. MacDonald, D.~K.
  Efetov,
  \href{https://www.nature.com/articles/s41586-019-1695-0}{Superconductors,
  orbital magnets and correlated states in magic-angle bilayer graphene},
  Nature 574~(7780) (2019) 653--657.

\bibitem{Serlin2020}
M.~Serlin, C.~L. Tschirhart, H.~Polshyn, Y.~Zhang, J.~Zhu, K.~Watanabe,
  T.~Taniguchi, L.~Balents, A.~F. Young,
  \href{https://www.science.org/doi/full/10.1126/science.aay5533}{Intrinsic
  quantized anomalous {Hall} effect in a moir{\'e} heterostructure}, Science
  367~(6480) (2020) 900--903.

\bibitem{Zhou2022}
H.~Zhou, L.~Holleis, Y.~Saito, L.~Cohen, W.~Huynh, C.~L. Patterson, F.~Yang,
  T.~Taniguchi, K.~Watanabe, A.~F. Young,
  \href{https://www.science.org/doi/full/10.1126/science.abm8386}{Isospin
  magnetism and spin-polarized superconductivity in {Bernal} bilayer graphene},
  Science 375~(6582) (2022) 774--778.

\bibitem{Geisenhof2021}
F.~R. Geisenhof, F.~Winterer, A.~M. Seiler, J.~Lenz, T.~Xu, F.~Zhang, R.~T.
  Weitz, \href{https://www.nature.com/articles/s41586-021-03849-w}{Quantum
  anomalous {Hall} octet driven by orbital magnetism in bilayer graphene},
  Nature 598~(7879) (2021) 53--58.

\bibitem{Ahn2018}
S.~J. Ahn, P.~Moon, T.-H. Kim, H.-W. Kim, H.-C. Shin, E.~H. Kim, H.~W. Cha,
  S.-J. Kahng, P.~Kim, M.~Koshino, Y.-W. Son, C.-W. Yang, J.~R. Ahn,
  \href{https://www.science.org/doi/full/10.1126/science.aar8412}{Dirac
  electrons in a dodecagonal graphene quasicrystal}, Science 361~(6404) (2018)
  782--786.

\bibitem{Yao2018}
W.~Yao, E.~Wang, C.~Bao, Y.~Zhang, K.~Zhang, K.~Bao, C.~K. Chan, C.~Chen,
  J.~Avila, M.~C. Asensio, J.~Zhu, S.~Zhou,
  \href{https://www.pnas.org/doi/abs/10.1073/pnas.1720865115}{Quasicrystalline
  30$^{\circ}$ twisted bilayer graphene as an incommensurate superlattice with
  strong interlayer coupling}, Proceedings of the National Academy of Sciences
  115~(27) (2018) 6928--6933.

\bibitem{Mele2010}
E.~J. Mele,
  \href{https://link.aps.org/doi/10.1103/PhysRevB.81.161405}{Commensuration and
  interlayer coherence in twisted bilayer graphene}, Physical Review B 81~(16)
  (2010) 161405.

\bibitem{Mele2012}
E.~J. Mele, \href{https://dx.doi.org/10.1088/0022-3727/45/15/154004}{Interlayer
  coupling in rotationally faulted multilayer gra\-phenes}, Journal of Physics D:
  Applied Physics 45~(15) (2012) 154004.

\bibitem{Mele2023a}
S.~Talkington, E.~J. Mele,
  \href{https://link.aps.org/doi/10.1103/PhysRevB.107.L041408}{Electric-field-tunable
  band gap in commensurate twisted bilayer graphene}, Physical Review B 107~(4)
  (2023) L041408.

\bibitem{Mele2023b}
S.~Talkington, E.~J. Mele,
  \href{https://link.aps.org/doi/10.1103/PhysRevB.108.085421}{Terahertz
  circular dichroism in commensurate twisted bilayer graphene}, Physical Review
  B 108~(8) (2023) 085421.

\bibitem{Kindermann2015}
M.~Kindermann,
  \href{https://link.aps.org/doi/10.1103/PhysRevLett.114.226802}{Topological
  {Crystalline} {Insulator} {Phase} in {Graphene} {Multilayers}}, Physical
  Review Letters 114~(22) (2015) 226802.

\bibitem{Koren2016}
E.~Koren, I.~Leven, E.~L{\"o}rtscher, A.~Knoll, O.~Hod, U.~Duerig,
  \href{https://www.nature.com/articles/nnano.2016.85}{Coherent commensurate
  electronic states at the interface between misoriented graphene layers},
  Nature Nanotechnology 11~(9) (2016) 752--757.

\bibitem{Riedl2010}
C.~Riedl, C.~Coletti, U.~Starke,
  \href{https://dx.doi.org/10.1088/0022-3727/43/37/374009}{Structural and
  electronic properties of epitaxial graphene on {SiC}(0001): a review of
  growth, characterization, transfer doping and hydrogen intercalation},
  Journal of Physics D: Applied Physics 43~(37) (2010) 374009.

\bibitem{Razado2018}
I.~Razado-Colambo, J.~Avila, D.~Vignaud, S.~Godey, X.~Wallart, D.~P. Woodruff,
  M.~C. Asensio,
  \href{https://www.nature.com/articles/s41598-018-28402-0}{Structural
  determination of bilayer graphene on {SiC}(0001) using synchrotron radiation
  photoelectron diffraction}, Scientific Reports 8~(1) (2018) 10190.

\bibitem{Kim2008}
K.-j. Kim, H.~Lee, J.-H. Choi, Y.-S. Youn, J.~Choi, H.~Lee, T.-H. Kang, M.~C.
  Jung, H.~J. Shin, H.-J. Lee, S.~Kim, B.~Kim,
  \href{https://onlinelibrary.wiley.com/doi/abs/10.1002/adma.200800742}{Scanning
  {Photoemission} {Microscopy} of {Graphene} {Sheets} on {SiO2}}, Advanced
  Materials 20~(19) (2008) 3589--3591.

\bibitem{deLima2013}
L.~H. de~Lima, A.~de~Siervo, R.~Landers, G.~A. Viana, A.~M.~B. Goncalves, R.~G.
  Lacerda, P.~H{\"a}berle,
  \href{https://link.aps.org/doi/10.1103/PhysRevB.87.081403}{Atomic surface
  structure of graphene and its buffer layer on {SiC}(0001): {A}
  chemical-specific photoelectron diffraction approach}, Physical Review B
  87~(8) (2013) 081403.

\bibitem{Matsui2013}
F.~Matsui, R.~Ishii, H.~Matsuda, M.~Morita, S.~Kitagawa, T.~Matsushita, S.~Koh,
  H.~Daimon,
  \href{https://iopscience.iop.org/article/10.7567/JJAP.52.110110/meta}{Characterizing
  {Edge} and {Stacking} {Structures} of {Exfoliated} {Graphene} by
  {Photoelectron} {Diffraction}}, Japanese Journal of Applied Physics 52~(11R)
  (2013) 110110.

\bibitem{Tricot2022}
S.~Tricot, T.~Jaouen, D.~S{\'e}billeau, P.~Schieffer,
  \href{https://www.sciencedirect.com/science/article/pii/S0368204822000172}{Energy
  dependence of interference phenomena in the forward-scattering regime of
  photoelectron diffraction}, Journal of Electron Spectroscopy and Related
  Phenomena 256 (2022) 147176.

\bibitem{Sebilleau2006}
D.~S\'ebilleau, R.~Gunnella, Z.-Y. Wu, S.~D. Matteo, C.~R. Natoli,
  \href{https://dx.doi.org/10.1088/0953-8984/18/9/R01}{Multiple-scattering
  approach with complex potential in the interpretation of electron and photon
  spectroscopies}, Journal of Physics: Condensed Matter 18~(9) (2006) R175.

\bibitem{Sebilleau2011}
D.~S\'ebilleau, C.~Natoli, G.~M. Gavaza, H.~Zhao, F.~Da~Pieve, K.~Hatada,
  \href{https://www.sciencedirect.com/science/article/pii/S0010465511002591}{{MsSpec}-1.0:
  {A} multiple scattering package for electron spectroscopies in material
  science}, Computer Physics Communications 182~(12) (2011) 2567--2579.

\bibitem{wwwMsSpec}
S.~Tricot, D.~S{\'e}billeau, \href{https://msspec.cnrs.fr}{{MsSpec} package,
  {Python} version}.

\bibitem{Hedin1970}
L.~Hedin, S.~Lundqvist,
  \href{https://www.sciencedirect.com/science/article/pii/S0081194708606153}{Effects
  of {Electron}-{Electron} and {Electron}-{Phonon} {Interactions} on the
  {One}-{Electron} {States} of {Solids}}, in: F.~Seitz, D.~Turnbull,
  H.~Ehrenreich (Eds.), Solid {State} {Physics}, Vol.~23, Academic Press, 1970,
  pp. 1--181.

\bibitem{Hedin1971}
L.~Hedin, B.~I. Lundqvist,
  \href{https://dx.doi.org/10.1088/0022-3719/4/14/022}{Explicit local
  exchange-correlation potentials}, Journal of Physics C: Solid State Physics
  4~(14) (1971) 2064.

\bibitem{Fujikawa2000}
T.~Fujikawa, K.~Hatada, L.~Hedin,
  \href{https://link.aps.org/doi/10.1103/PhysRevB.62.5387}{Self-consistent
  optical potential for atoms in solids at intermediate and high energies},
  Physical Review B 62~(9) (2000) 5387--5398.

\bibitem{Kuttel1994}
O.~M. K{\"u}ttel, R.~G. Agostino, R.~Fasel, J.~Osterwalder, L.~Schlapbach,
  \href{https://www.sciencedirect.com/science/article/pii/0039602894908109}{X-ray
  photoelectron and {Auger} electron diffraction study of diamond and graphite
  surfaces}, Surface Science 312~(1) (1994) 131--142.

\bibitem{Maillard1996}
E.~Maillard-Schaller, O.~M. Kuettel, L.~Schlapbach,
  \href{https://onlinelibrary.wiley.com/doi/abs/10.1002/pssa.2211530216}{X-ray
  photoelectron diffraction of the silicon–diamond interface}, physica status
  solidi (a) 153~(2) (1996) 415--429.

\bibitem{Greber1998}
T.~Greber, J.~Wider, E.~Wetli, J.~Osterwalder,
  \href{https://link.aps.org/doi/10.1103/PhysRevLett.81.1654}{X-{Ray}
  {Photoelectron} {Diffraction} in the {Backscattering} {Geometry}: {A} {Key}
  to {Adsorption} {Sites} and {Bond} {Lengths} at {Surfaces}}, Physical Review
  Letters 81~(8) (1998) 1654--1657.

\bibitem{Auwarter1999}
W.~Auw{\"a}rter, T.~J. Kreutz, T.~Greber, J.~Osterwalder,
  \href{https://www.sciencedirect.com/science/article/pii/S0039602899003817}{{XPD}
  and {STM} investigation of hexagonal boron nitride on {Ni}(111)}, Surface
  Science 429~(1) (1999) 229--236.

\bibitem{Matsui2012}
F.~Matsui, T.~Matsushita, H.~Daimon,
  \href{https://journals.jps.jp/doi/10.1143/JPSJ.81.114604}{Photoelectron
  {Diffraction} and {Holographic} {Reconstruction} of {Graphite}}, Journal of
  the Physical Society of Japan 81~(11) (2012) 114604.

\bibitem{Roth2013}
S.~Roth, F.~Matsui, T.~Greber, J.~Osterwalder,
  \href{https://doi.org/10.1021/nl400815w}{Chemical {Vapor} {Deposition} and
  {Characterization} of {Aligned} and {Incommensurate} {Graphene} / {Hexagonal}
  {Boron} {Nitride} {Heterostack} on {Cu}(111)}, Nano Letters 13~(6) (2013)
  2668--2675.

\bibitem{Agliz1995}
D.~Agliz, A.~Qu{\'e}merais, D.~S{\'e}billeau,
  \href{https://www.sciencedirect.com/science/article/pii/0039602895008152}{Splitting
  effects in high resolution and high energy photoelectron diffraction: the
  case of {MgO}(001)}, Surface Science 343~(1) (1995) 80--86.

\bibitem{Juillaguet1995}
S.~Juillaguet, L.~Kubler, M.~Diani, J.~L. Bischoff, G.~Gewinner, P.~Wetzel,
  N.~B{\'e}court,
  \href{https://www.sciencedirect.com/science/article/pii/0039602895006761}{Strong
  element dependence of {C} 1s and {Si} 2p {X}-ray photoelectron diffraction
  profiles for identical {C} and {Si} local geometries in $\beta$-{SiC}},
  Surface Science 339~(3) (1995) 363--371.

\bibitem{Schieffer2001}
P.~Schieffer, G.~J{\'e}z{\'e}quel, B.~L{\'e}pine, D.~S{\'e}billeau,
  G.~Feuillet, B.~Daudin,
  \href{https://www.sciencedirect.com/science/article/pii/S0039602800010499}{X-ray
  photoelectron diffraction from cubic {GaN}(001): an experimental and
  theoretical study}, Surface Science 482-485 (2001) 593--599.

\bibitem{Sadoc2023}
J.-F. Sadoc, M.~Imp{\'e}ror-Clerc,
  \href{https://dx.doi.org/10.1209/0295-5075/ad16f3}{Some examples of
  quasiperiodic tilings obtained with a simple grid method}, Europhysics
  Letters 144~(6) (2024) 66002.

\bibitem{Woodruff1992}
V.~Fritzsche, D.~P. Woodruff,
  \href{https://link.aps.org/doi/10.1103/PhysRevB.46.16128}{Direct
  photoelectron-diffraction method for adsorbate structural determinations},
  Physical Review B 46~(24) (1992) 16128--16134.

\bibitem{Woodruff1994}
D.~P. Woodruff, A.~M. Bradshaw,
  \href{https://dx.doi.org/10.1088/0034-4885/57/10/003}{Adsorbate structure
  determination on surfaces using photoelectron diffraction}, Reports on
  Progress in Physics 57~(10) (1994) 1029.

\end{thebibliography}

\end{document}